\documentclass[showpacs,showkeys,floatfix]{revtex4}
\usepackage[dvips]{graphicx}
\usepackage{epsfig}

\usepackage[cp1251]{inputenc}
\usepackage[english, russian]{babel}
\usepackage{color}
\usepackage{ulem}
\usepackage{bm}

\begin{document}

\title{Twisted quasiperiodic textures of biaxial nematics.}

\author{V.L. Golo} 
\affiliation{Department of Mechanics and Mathematics, Lomonosov Moscow
State University\\ Moscow, Russia, and \\
National Research University Higher School of Economics\\
Moscow, Russia }

\author{E.I. Kats}
\affiliation{Landau Institute for Theoretical Physics \\
Chernogolvka, Moscow region, Russia}    
\email{kats@landau.ac.ru}

\author{A.A. Sevenyuk}  
\affiliation{Department of Mechanics and Mathematics,  Lomonosov Moscow
State University\\ Moscow, Russia}

\author{D.O. Sinitsyn} 
\affiliation{Semenov Institute of Chemical Physics\\
Moscow, Russia}

\begin{abstract}
Textures (i.e., smooth space non-uniform distributions of the order parameter) in biaxial nematics turned out to be much more
difficult and interesting than expected. Scanning the literature we find only a very few publications on this topic.
Thus, the immediate motivation of the present paper is to develop a systematic procedure to study, classify and visualize
possible textures in biaxial nematics.  Based on the elastic energy of a biaxial nematic (written in the most simple form
that involves the least number of phenomenological parameters) we derive and solve numerically 
the Lagrange equations of the first kind. It allows one to visualize the solutions
and offers a deep insight into their geometrical and topological features. Performing Fourier analysis we find some particular 
textures possessing two
or more characteristic space periods (we term such solutions quasiperiodic ones because the periods are not necessarily commensurate).
The problem is not only of intellectual interest but also of relevance to
optical characteristics of the liquid-crystalline textures. 
\end{abstract}



\maketitle

\section{Introduction. }
\label{int}
The combination of orientational order (leading to non-trivial optical response) and
relatively soft elasticity (leading to high sensitivity to boundary conditions) makes liquid crystals very interesting
systems possessing rich variety of textures (topologically stable 
smooth configurations of the order parameter).
Although textures and related optical properties of uniaxial nematics are well studied and results of the investigations have numerous applications,
there are still only a few publications on textures in biaxial nematics, i.e. when molecular orientations are arranged regularly 
in two directions (see e.g. \cite{CP05}, \cite{KV09}). In this work we investigate geometrical and topological
features of textures in biaxial  liquid crystals. We show that there are some particular configurations with quasiperiodic
(i.e., possessing two or more generally speaking incommensurate periods) space distribution of the order parameter.

In what follows we always have in mind a liquid crystalline slab placed between two polarizers.
The nematic order parameter (director ${\bf n}$ for uniaxial nematics  or three mutually orthogonal
unit vectors ${\bf n}$, ${\bf l}$, and ${\bf m} = {\bf n} \times {\bf l}$ for biaxial nematics) is uniform in-plane (independent of
$x$ and $y$) \cite{GP93}, \cite{PI91}. 
However a question of how to identify biaxial nematics is still a problem. One evident approach - optical microscopy- would be to examine the optical properties of the
textures between crossed polarizers.
Dealing with optical properties we assume that the light beam is propagating along the normal to the slab ($z$-axis). In such a situation for uniaxial
nematics one can learn almost everything about coarse grained features of the textures from polarizing microscopy 
investigations which give average 2D information integrated 
over the path of the light (or to get full 3D information more sophisticated and expensive confocal
polarizing microscopy methods \cite{SL10} should be used). Indeed the pattern of transmitted light intensity depends on the difference
of times for ordinary and extraordinary beams to pass through the slab. If the uniaxial director is everywhere parallel to $z$-axis, there is
no transmitted light at all in the crossed polarizers. Therefore non-zero intensity indicates deviations from uniform
alignment (${\bf n} = (0\, , 0\, , 1)$). 

Behavior is so simple because for the uniaxial nematics the optical axis is uniquely
defined by the director (either parallel to the director for prolate permeability tensor, or perpendicular to it
for the oblate tensor). Even more, all physical property tensors contain elements expressed in terms of laboratory axes,
and the axes are simply related to crystallographic Cartesian axes.
Everything is not so simple for biaxial nematics, where all three eigenvalues
of the order parameter tensor (i.e., dielectric permeability tensor) are different.
Although the system of Maxwell's equations irrespective to the liquid crystal symmetry (uniaxial or biaxial)
can be always described in terms of only two normal modes, the transmitted intensity for biaxial textures can be
zero only for two special configurations, corresponding to the light propagating along so-called biradials \cite{LL63} (see more details
in \cite{MR93}).
In their turn the biradial directions depend on the biaxiality and on three unit vectors ${\bf n}$, ${\bf l}$, ${\bf m}$.
These quantities for a generic biaxial nematic texture are  
non-trivially oriented with respect to the light propagation direction and distributed over $z$.
We conclude that in biaxial nematics there are intrinsic relations between geometry and topology of the order parameter textures
and optical properties expressed in terms of geometry and topology of the wave or ray surfaces (determined
by Fresnel's equations \cite{LL63}). It must be frankly admitted that we know practically nothing about optical properties of
biaxial nematic textures, and the first step to get insight on optical properties is to calculate
and then visualize the biaxial order parameter textures. 
We believe that not only the study of biaxial nematic textures is an interesting problem in its own right but as well 
the basic ideas inspiring our work can be applied to a large variety
of other interesting problems of textures in liquid crystals.

The phenomenological model and the order parameters on which our description is based are introduced in the next section
\ref{biax}. Technical details of our variational approach and computation method are presented in section \ref{var}.
The main results of our work are summarized in section \ref{res}. In the conclusion section \ref{con} we discuss significance and limitations
of our findings.

\section{Biaxial nematic: Background.}
\label{biax}
Biaxial nematics ($N_b$) possess three soft (Goldstone) orientational degrees of freedom: two corresponding to the direction 
of the long axis of a molecule and one degree for the rotations of the molecule around this axis. 
Thus, they have the same freedom of orientations in space as a solid body. Consequently, they should have three optical axes,
and, in comparison with uniaxial nematics, an additional pair of diffuse (liquid-like) X-ray diffraction peaks.
It should be noted that biaxial nematics have been experimentally identified, see \cite{saupe} - \cite{chan3} and also more recent 
publications \cite{DM06}, \cite{VP08}, \cite{PD12};
an important tool for their investigation being the X-ray scattering, \cite{feiser}, \cite{windle}.
The symmetry structure of biaxial nematics requires the use of a matrix order parameter within the framework of the 
Landau-de Gennes theory, \cite{GP93} - \cite{PI91}. 
The guidelines to the effect are due to the theory
of uniaxial nematics ($N_u$).  It was de Gennes who suggested that the order parameter for $N_u$ phase may be cast in the form
\begin{equation}
   \label{eq:nemdegennes}
   Q_{ij} = S_u \cdot \left(   n_i n_j   - \frac{1}{3} \delta_{ij} \right)
\end{equation}
where $S_u$ is the module of the order parameter, whereas $n_i$ are coordinates of the director, 
that is a vector indicating the average orientation of molecules. 
Similarly, the orientation in biaxial nematics relies on the use of a frame structure,
that is the order parameter of the form, which is a generalization of that given by Eq.(\ref{eq:nemdegennes}),
\begin{equation} 
    \label{eq:two_vec}
    Q_{ij} = S_u ( n_i n_j - \frac{1}{3} \delta_{ij} ) + S_b ( m_i m_j - l_i l_j),
\end{equation}
where $S_b$ describes the system's biaxiality.    
Biaxial nematics are studied by employing this frame.

In this paper we shall use a method that is equivalent to the above approach, but provides useful tools for studying spatial inhomogeneities
of the orientational order, or {\it textures}.  It is to be noted that Eq.(\ref{eq:two_vec}) indicates that the matrix of the order parameter
${\hat Q}$
may be cast in the form
\begin{equation}
   \label{eq:order}
       {\hat Q} = {\hat R}^{-1} {\hat Q}_0 {\hat R}
\end{equation}
where ${\hat R}$ is a rotational $3 \times 3 $ matrix and matrix ${\hat Q}_0$ reads
\begin{equation}
   {\hat Q}_0 =\left( 
	 \begin{array}{ccc}
	        \lambda_1 &   0          &  0                              \\
		 0         &   \lambda_2 &  0                              \\
		 0         &  0          &  - ( \lambda_1   +  \lambda_2 ) \\
         \end{array}
        \right)   
\label{eq:init}
\end{equation}
The representation (\ref{eq:order}) of the order parameter 
opens a convenient way for the application of topology to various problems 
of field theory and condensed matter, \cite{monst2}.  Recently, Monastyrskii and Sasorov, \cite{monst3}, 
employed the method for studying singularities in biaxial nematics.  
It is important that values of the order parameter may change from one region of the volume to another.
In this sense they are local characteristics of the system's state.

The key point about the study of the phenomena is to employ
the continuum theory by considering the order parameter as a variable that {\it describes the intrinsic structure of the liquid}, \cite{GP93}.
It can be incorporated in the free energy through appropriate terms that may also describe its spatial variations,
and one can employ the minimization of the free energy  for writing down the equation of textures.  
To that end we shall write down the part of the free energy related to spatial inhomogeneities in the form 
$$
  {\cal F}_{\nabla} = \int F_{\nabla} d^3 x
$$
where the density of the free energy reads
\begin{equation}
   \label{eq:grad}
   F_{\nabla} = K_1 \partial_i Q_{ki} \partial_j Q_{kj} +  K_2 \partial_i Q_{kj} \partial_i Q_{kj} + K_3 \partial_i Q_{kj} \partial_j Q_{ki} 
\end{equation}
It is worth noting that we have little knowledge as to the relative size of the coefficients $K_1,K_2,K_3$  for the biaxial nematics.  
It is equally important that, applying integration by parts to the bulk free energy, we may substantially change the form of $F_{\nabla}$ 
through taking into account appropriate boundary conditions.
The minimization equations for $F_\nabla$ given by Eq.(\ref{eq:grad}) are extremely hard to solve analytically. 
Therefore we consider a special but very important case of one dimensional textures. 
Thus we assume that the order parameter depends only on one spatial coordinate, say $z$. 
This approximation can describe textures arising in the commonly used experimental configuration
in which the liquid crystal is placed in a thin layer between two planar uniform glass substrates.

\section{Variational principle: Technical details. }
\label{var}
\noindent
The gradient part of the free energy for one dimensional textures follows from Eq.(\ref{eq:grad}) by neglecting terms with derivatives in $x_1= x, x_2=y$,
and preserving only terms in $x_3 = z$.  In what follows we shall use the notation
$$
     \frac{d}{dz} f = \dot f
$$
for any function $f$ of z. Thus we obtain the following expression for the gradient energy
\begin{equation}
      \label{eq:gradone}
      F_{\nabla} = K_2 Tr \left( 
                                {\hat {\dot Q}} \cdot    {\hat {\dot Q}} 
			  \right)
                          +  ( K_1 + K_3) \left( 
			                         {\hat {\dot Q}} \cdot   {\hat {\dot Q}}  
				          \right)_{33}
\end{equation}
We have to minimize the functional of the gradient part of the free energy
\begin{equation}
      \label{eq:functional}
      {\cal F} = \int \limits_0^z \, F_{\nabla} \, dz
\end{equation}
under the constraint that the symmetric matrix 
of the order parameter have fixed eigenvalues, see Eq.(\ref{eq:order}), and trace zero.  
This requirement may be accommodated by imposing the constraints
\begin{equation}
   \label{eq:constr}
    Tr({\hat Q}) =0; \quad Tr({\hat Q} \cdot {\hat Q}) = const_1; \quad Tr({\hat Q} \cdot {\hat Q} \cdot {\hat Q}) = const_2
\end{equation}
The minimization problem may be solved either by the use of the Lagrangian equations of the 1-st kind, that is using Lagrangian multipliers, 
or by resolving the above constraints and employing the Lagrangian equations of the 2-nd kind. To that end we shall use the representation
given by Eq.(\ref{eq:order}).

The Lagrangian equations of the 1-st kind are more appropriate for numerical simulation.  Generally, it is a hard problem to resolve constraints.
It is equally important that the constraints being resolved, the obtained equations depend on the choice of local parameters, that is generalized coordinates,
and therefore subject to some non-evident tricks. For that reason we shall use the equations of the 1-st kind for our numerics,  and  introduce the effective 
Lagrangian
\begin{equation}
      \label{eq:efflagr}
      {\cal L} = F_{\nabla} - \Lambda_1 \; Tr({\hat Q}) - \Lambda_2 \; Tr({\hat Q} \cdot {\hat Q})   - \Lambda_3 \; Tr({\hat Q} \cdot {\hat Q} \cdot 
      {\hat Q}) 
\end{equation}
and solve the minimization problem in accord with the usual rules. It should be noted that we find only extremal solutions, which, generally speaking, minimize the functional only for sufficiently short intervals in $z$.

\noindent
To describe the change of position of a molecule in space we may employ the skew-symmetric matrix
\begin{equation}
    \label{eq:angmatrix}
   {\hat \Omega } = {\hat R}^{-1} \dot {\hat R}
\end{equation}
where ${\hat R}$ is the rotation matrix defining the value of the order parameter, see Eq.(\ref{eq:order}). Matrix ${\hat \Omega } $ is intimately related
to the usual vector of angular velocity, $\bm \omega$, through the equation
$$
    \Omega_{ij} = - \epsilon_{ijk} \, \omega_k
$$
or in the matrix form
$$
   {\hat \Omega } = \sum \limits_{k=1}^3  \omega_k \, f^k
$$
where the $f^k$ are generators of rotations about axes $\hat x_k, \; k=1,2,3$,
and the components $\omega_k$ of the angular velocity vector can be expressed in terms of the Euler angles
$\theta, \varphi, \psi$ for the rotation matrix ${\hat R}$ by the standard formulas, see~\cite{LL78}:
$$
\begin{array}{lcl}
\omega_1 = \dot \varphi \sin{\theta} \sin{\psi} + \dot \theta \cos{\psi}, \\
\omega_2 = \dot \varphi \sin{\theta} \cos{\psi} - \dot \theta \sin{\psi}, \\
\omega_3 = \dot \varphi \cos{\theta} + \dot \psi.
\end{array}
$$

\section{Equations of the textures}
\label{eqs}

As explained above, for the numerical analysis, we employ the Lagrange equations of the first kind
derived from the Lagrangian (\ref{eq:efflagr}),(\ref{eq:gradone}), which have the general form:
$$
\frac{d}{dt} \frac{\partial \cal L}{\partial \dot Q_{ij}} =
\frac{\partial \cal L}{\partial Q_{ij}}.
$$
These equations contain the
Lagrange multipliers $\Lambda_1, \Lambda_2, \Lambda_3$, which we express in terms of the
elements of ${\hat Q}\, , {\hat {\dot Q}}$ using the constraints (\ref{eq:constr})
according to the standard procedure of Lagrange multipliers exclusion (see e.g. \cite{Gantm}).
The expressions used along the way get quite involved very quickly,
so an essential component of success in the work with systems of such analytical complexity
is the automation of the work with formulas provided by means of symbolic computation.

After the explicit equations are obtained, approximate solutions can be found numerically.
For describing the possible textures of the system we numerically solve the Cauchy
problem for the Lagrange equations with
a diagonal initial order parameter matrix ${\hat Q}(0)$ given by (\ref{eq:init})
and a given initial derivative matrix ${\hat {\dot Q}}(0)$,
which we compute using the representation (\ref{eq:order})
and choosing an arbitrary initial angular velocity vector $\bm\omega(0)$.
This guarantees the tangency of the initial velocity to the manifold defined by
the constraints (\ref{eq:constr}).

From here on we use the following notation for the parameters of the Lagrangian:
$$
k_1 = K_2, \quad k_2 = K_1 + K_3.
$$

It is also worthwhile to note that the terms ``evolution'', ``velocity'',
the ``dot'' derivatives {\it etc.} refer in our context to the change of the variables along the Z axis.

\section{Results of the simulations}
\label{res}

\subsection{Homogeneous phase}

The simplest and trivial arrangement of the orientational order parameters arises in the case of zero initial velocity, in which the
order parameter is constant (for any values of $k_1, k_2$), and the structure is homogeneous. 
Correspondingly in this case all biaxial molecules (which can be visualized as identical bricks) have the same orientation.

\subsection{Simple Helicoidal phase}

\begin{figure}
  \begin{center}
    \includegraphics[height = 200bp]{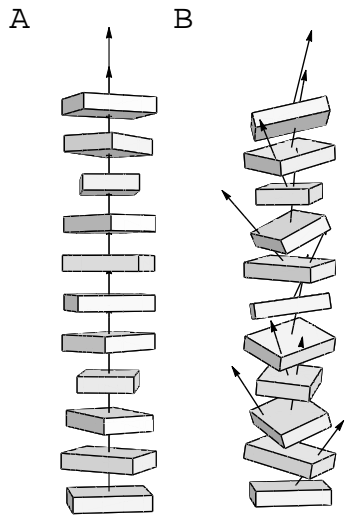}
    \caption{Arrangement of molecules. A: Helicoidal phase, $\bm\omega(0) = (0, 0, 1)$,
	B: Quasiperiodic phase, $\bm\omega(0) = (\frac13, \frac12, 1)$. Arrows indicate the angular velocity vectors. Other parameters in both pictures have the values:
	$k_1 = 1, k_2 = \frac12, \lambda_1 = 7, \lambda_2 = -10$.}
    \label{bricks}
  \end{center}
\end{figure}

The Lagrange equations for the order parameter (for any values of $k_1, k_2$) admit of a partial exact solution of the form:
\begin{equation}
{\hat Q}(z) = {\hat R}_3^{-1}(\omega z) {\hat Q}(0) {\hat {R}_3(\omega z)},
\label{cholSolution}
\end{equation}
where ${\hat R}_3(\omega z)$ is the matrix of the three-dimensional rotation about the Z axis through the angle $\omega z$, and ${\hat Q}(0)$ is given by (\ref{eq:init}).
The fact that (\ref{cholSolution}) satisfies the Lagrange equations was checked using
symbolic computation. The angular velocity for this solution does not depend on z:
$$
{\bm \omega } = (0,0,\omega).
$$
The eigenvectors of the matrix ${\hat Q}(z)$ for this solution have the form:
$$
\begin{array}{lcl}
{\bf l} = (-\cos{\omega z}, \sin{\omega z}, 0), \\
{\bf m} = (\sin{\omega z}, \cos{\omega z}, 0), \\
{\bf n} = (0, 0, 1).
\end{array}
$$
Thus, from layer to layer, the directions of the molecule's axes perform uniform rotation in the XY plane
with the angular velocity $\omega$.
The arrangement of molecules corresponding to this solution is shown
in Fig.~\ref{bricks},~A. This phase shares some characteristics with the {\it cholesteric} structure,
the evolution of the molecule orientation from layer to layer consisting in
the rotation about the Z axis with the angular velocity $\omega $. 
The period of this structure in z (also known as the {\it pitch}) has the value
$T = {\pi}/{\omega}$
($\pi $ and not $2\pi $ because in non-polar biaxial nematics ${\bf n}$, ${\bf m}$, and ${\bf l}$ are
physically equivalent to ${-\bf n}$, ${-\bf m}$, and ${-\bf l}$ respectively, thus
a rotation through $\pi$ radians transforms an order parameter matrix into itself).
This similarity is a natural consequence of a close connection between chirality (mirror symmetry breaking)
and biaxiality (i.e., additional to uniaxial director ${\bf n}$ ordering in the perpendicular to ${\bf n}$ plane).
In our case chirality is produced externally by applied on the boundary twist ${\hat \Omega }$ and biaxiality
is an intrinsic property of the liquid crystalline material (therefore can be arbitrarily large). In conventional cholesterics
(see e.g. \cite{HK99}) chirality is a material property, whereas biaxiality is a geometrical consequence of simple
spiral twist structure and is typically very small.

\subsection{Generic quasiperiodic textures}

When the initial angular velocity direction differs from the Z axis, a more complicated
solution arises. The molecular arrangement of this type is shown in
Fig.~\ref{bricks},~B. One can see that the angular velocity is changing
from layer to layer, and the molecular orientation planes are not parallel.

\begin{figure}
  \begin{center}
    \includegraphics[height = 150bp]{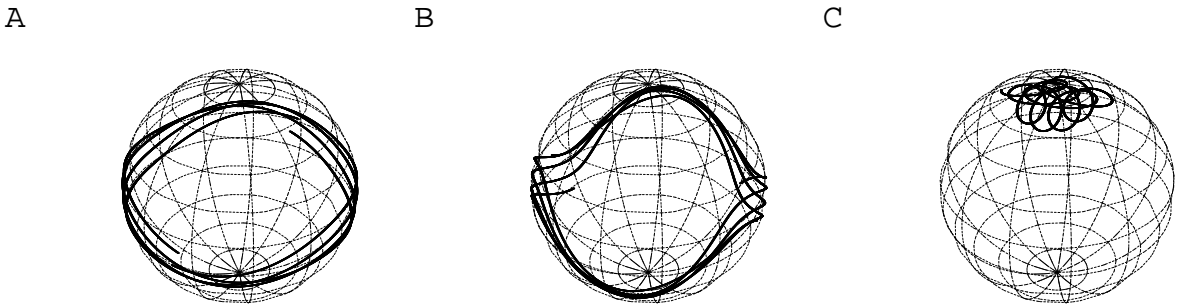}
    \caption{A,B,C: Trajectories of the eigenvectors $\bf l\, , \bf m\, , \bf n$
	(correspondingly) of the order parameter matrix in the quasy-periodic phase.
	The parameters have the values: $\bm\omega(0) = (\frac13, \frac12, 1),
	k_1 = 1, k_2 = \frac12, \lambda_1 = 7, \lambda_2 = -10$.}
    \label{vecs}
  \end{center}
\end{figure}

To study this structure in more detail, we draw the trajectories of the eigenvectors
$\bf l\, , \bf m\, , \bf n$ of the order parameter matrix ${\hat Q}$, which correspond to the directions
of the molecules' axes, see Fig.~\ref{vecs}. In the case illustrated
the vector $\bf m$ corresponds to the eigenvalue with the largest magnitute,
depicted in Fig.~\ref{bricks},~B by the longest
axis of the parallelepiped representing the molecule.
Its trajectory is shown in Fig.~\ref{vecs},~B. One can see the rotation
of this vector about the Z axis (as in the helicoidal phase) combined with an oscillatory motion in the vertical direction.
Thus, in contrast to the helicoidal phase, this texture has more than one characteristic
frequency, exhibiting quasiperiodic structure.
The third eigenvector $\bf n$ corresponds in Fig.~\ref{bricks} to
the shortest axis of a parallelepiped, or, in other words, the normal to the plane of the molecule.
The evolution of this normal is shown in Fig.~\ref{vecs},~C. 
In the helicoidal structure this vector is directed along the Z axis,
whereas in the quasiperiodic structure the Z direction is only the average
position of $\bf n$, while its momentary values oscillate around this mean orientation.

\begin{figure}
  \begin{center}
    \includegraphics[width = 450bp]{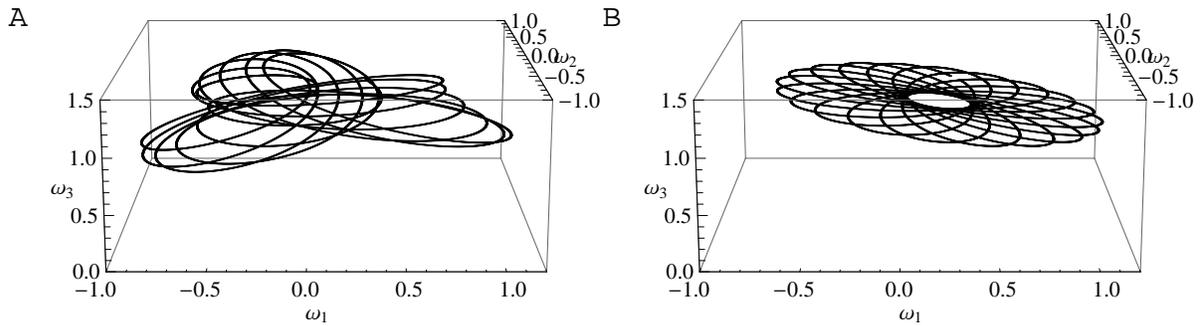}
    \caption{Trajectory of the angular velocity. A: $k_2 = \frac12$, B: $k_2 = 0$.
	Other parameters in both pictures have the values:
	$\bm \omega(0) = (\frac13, \frac12, 1), k_1 = 1, \lambda_1 = 7, \lambda_2 = -10$.}
    \label{omega}
  \end{center}
\end{figure}

\begin{figure}
  \begin{center}
    \includegraphics[width = 250bp]{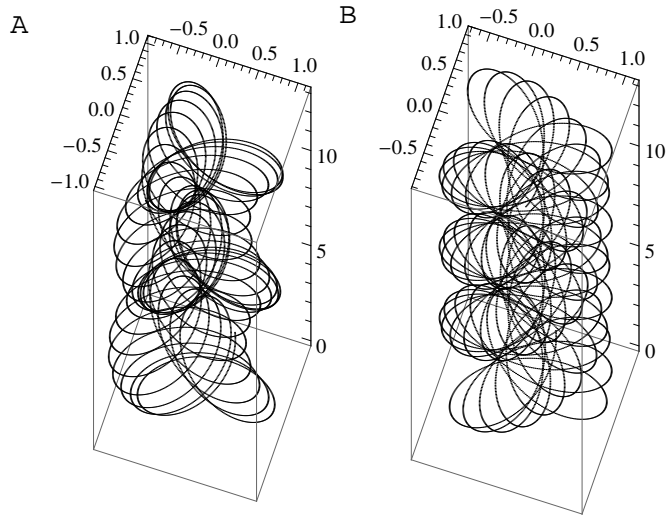}
    \caption{Curve parametrically defined as $(0\, ,0\, , z) + {\bm \omega }(z)$
	\ -- the curve traced by the end of the angular velocity vector drawn from the location of the corresponding
	molecule in the space.
	A: $k_2 = \frac12$, B: $k_2 = 0$.
	Other parameters in both pictures have the values:
	$\bm \omega(0) = (\frac13, \frac12, 1), k_1 = 1, \lambda_1 = 7, \lambda_2 = -10$.}
    \label{omegaCurve}
  \end{center}
\end{figure}

To throw more light upon the change of the orientational order along $Z$ axis, we also
investigate the evolution of the angular velocity vector for the quasiperiodic solution.
Its trajectory is shown in Fig.~\ref{omega},~A. In the particular case considered,
the 3rd component of the angular velocity doesn't change its sign, while the other two
components oscillate around zero. This agrees with Fig.~\ref{bricks},
where we see the same main pattern as in the helicoidal phase, that is the rotation of molecules
about the Z axis from layer to layer, but with an additional pattern governing the
precession of the molecular  planes. Another picture illustrating this structure is shown in
Fig.~\ref{omegaCurve}, where the ends of the angular velocity vectors are depicted, when each
vector is drawn from the location of the corresponding molecule,
i.e. the figure shows the vector $(0\, ,0\, , z) + {\bm \omega }(z)$.
A segment of this curve resembles the trefoil knot. This confirms that the structure has
more than one characteristic frequency.
A similar curve appears also in the case $k_2=0$, Fig.~\ref{omega},~B, Fig.~\ref{omegaCurve},~B.
In this case there are additional first integrals corresponding to the rotational symmetry
of the Lagrangian. This property affects the shape of the solution, which nonetheless retains
its quasiperiodic character.

\subsection{Spectral properties of the textures.}

\begin{figure}
  \begin{center}
    \includegraphics[width = 400bp]{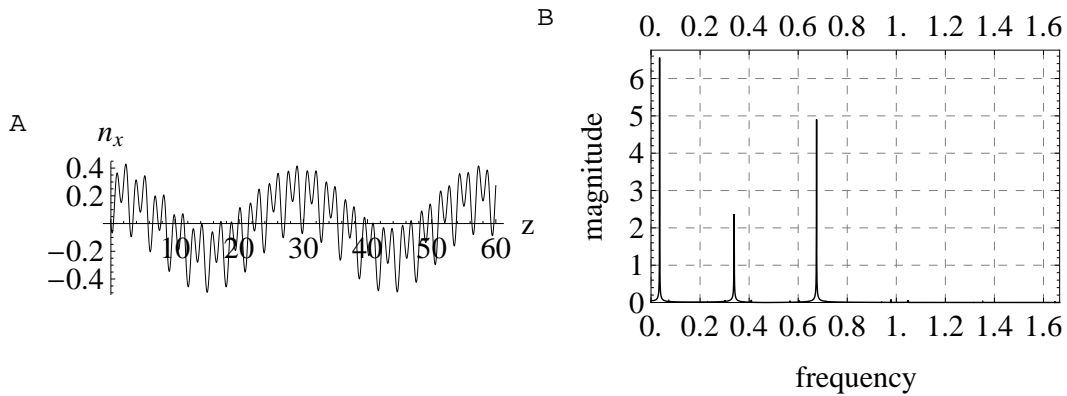}
    \caption{A: The X component of the third eigenvector $\bf n$ of the order
	parameter matrix as a function of $z$ in the quasiperiodic case. B: Spectrum of this function.
	One can see three peaks in the spectrum corresponding to three characteristic frequencies of the
	quasiperiodic texture. 
	The parameters have the values: $\bm\omega(0) = (\frac13, \frac12, 1),
	k_1 = 1, k_2 = \frac12, \lambda_1 = 7, \lambda_2 = -10$.}
    \label{spec}
  \end{center}
\end{figure}

The characteristic frequencies (in $z$) of the orientation patterns
and the corresponding (generalized) pitches,
i.e. the lengths at which the structure approximately repeats itself, may indicate the wavelengths
of light that correspond to special optical properties of these textures.
It is thus important to study the spectral characteristics of the solutions,
sampling the functions of $z$ with a sufficiently small step in an appropriate interval
and computing the discrete Fourier transform (the spectrum).
As a characteristic quantity we consider the X component $n_x$ of the third eigenvector ${\bf n}$ of
the order parameter matrix. Fig.~\ref{spec} shows this function and a plot of the magnitude of its discrete
Fourier transform.
One can see three major peaks in the spectrum corresponding to three characteristic frequencies
of the texture. This confirms the quasiperiodic nature of the texture under consideration
and indicates the frequencies of light at which special optical properties of the texture
may be expected.

The general structure of the set of characteristic frequencies and their dependence
on the parameters of the system needs further exploration.
The observed peaks may correspond to
linear combinations of a number of basic frequencies,
as is the case in the spectra of quasiperiodic functions.
One of the situations in which quasiperiodic solutions can appear
is the case of integrable, or close to integrable, Lagrange equations.
Thus, the nature of the characteristic frequencies
in this system may be clarified by a further analysis of its dynamical
properties, including the study of the first integrals and possible dimensionality reductions.

\section{Conclusions.}
\label{con}
In this paper we have presented a theory for orientational textures in biaxial nematic liquid crystals.
The theory combines two parts: the Landau - de Gennes free energy expansion over the gradients of the tensor order
parameter, and numeric solution of the Lagrange equations of the first kind.
The deceptively simple general main message of our work is that description of biaxial nematic textures involves new symmetry 
and topology features, admit the Hamiltonian formulation, and could result in new optical phenomena. In particular, 
in this paper we develop a systematic numeric procedure to study, classify and visualize
possible textures in biaxial nematics.  Based on the elastic energy of a biaxial nematic written in the most simple form
we derive and solve numerically the Lagrange equations of the first kind.
Performing Fourier analysis for the biaxial order parameter texture we find some particular 
textures possessing three characteristic space periods (which are not necessarily commensurate). Such a multi-periodic twisted
structure can be considered as a sort of artificial photonic crystal. One can expect very unusual optical characteristics,
e.g., two or three band gaps (unlike one band gap for single periodic cholesteric photonic crystals \cite{KA71}, \cite{BD79}).
Our results can be verified either in particle-resolved computer simulations or in optic experiments.
Future work should extend the present study to three spatial dimensions of the textures, which would require
more heavy and sophisticated numerical investigations but promises a reacher plethora of structures
interesting for optical applications.

The results of the present paper allow us to bring forward some hypotheses concerning other physical properties of biaxial nematics. 
The fact is that the three directors ${\bf n}\, , {\bf m}\, , {\bf l}$ -- the main axes of the orientational
order parameter -- are not necessarily the principal axes for all macroscopic second order tensorial physical properties. Such requirement is
neither warranted by a theory, nor imposed by experimental data. A variety of different symmetries are possible depending on the additional symmetry operations for the material under consideration. 
Our guess relies mainly on multiperiodical structure of their textures, which was discussed above.
For example the multi-periodic textures found in our work should result in additional pairs of X-ray diffraction peaks.
As a note of caution we should also mention that for the optical or X-ray methods
with a response integrated over the sample thickness, conventional uniaxial nematic textures (where the single optic axis varies in the three-dimensional space) can be confused with a response from a genuine biaxial nematic with two optical axes.
The methods of this paper could be extended to other systems, for example, DNA solutions.
Under appropriate conditions there exist  DNA phases  that  are believed to be cholesteric liquid crystals. 
The phases are prepared from segments of the DNA of a size  approximately equal to the persistence length, or $50 \, nm$. It is important 
that the segments do not have the rotational symmetry about their axes owing to the structure of the double helix.
Therefore, one may expect their being strongly biaxial (unlike only weakly biaxial conventional cholesterics
formed by molecules of low molecular mass), see \cite{JETPlett}.

The theory presented in our work can also have implications not only for optical or X-ray experiment qualitative rationalization.
One might think about various electro-optic and magneto-optic applications of biaxial nematic liquid crystals.
For example the response of the ''biaxial'' directors ${\bf m}\, , {\bf l}$  can be much faster than that of the uniaxial
director ${\bf n}$.

\acknowledgements

V.L.G. acknowledges the support of the Program Progress of Basic Research,
National Research University Higher School of Economics.

E.K. acknowledges the support of the RFBR grant No 13-02-00120 and hospitality of the Issac Newton Institute
for Mathematical Science.

D.S. acknowledges the support of the Government of the Russian Federation grant 
for support of research projects implemented by leading scientists at Lomonosov Moscow State University 
under the agreement  No. 11.G34.31.0054.

\end{document}